\documentstyle[prl,aps]{revtex}

\tighten

\begin{document}

\twocolumn[
\title{\bf Simulations of deposition growth models in various
dimensions.\\
Are overhangs important?}
\author{David Y.K. Ko and Flavio Seno}
\address{Department of Physics, University of Oxford, 1
Keble Road, Oxford, OX1 3NP, U.K.}

\widetext \leftskip=0.10753\textwidth
\rightskip\leftskip\begin{abstract}
We present simulation results of deposition growth of surfaces in
$2$\/, $3$\/  and $4$\/  dimensions for
ballistic deposition where overhangs are present, and for
restricted solid on solid deposition where there are no overhangs.
The values of the scaling exponents for the two models are found to be
different,
suggesting that they belong to different
universality classes.
\bgroup\draft\egroup
\end{abstract}
\maketitle]

The deposition growth of surfaces \cite{Viscek.and.Family}
has been a subject
of long continual theoretical and experimental interest \cite{interest}
due to its relevance to
non-equilibrium processes in general as well as its possible role in
surface technology.
The profile of the deposited
surface gradually roughens
under the stochastic accumulation of particles, and early simulations
by Family {\it et. al.}\/ \cite{Family} suggested that the surface
roughness exhibits a dynamical scaling behaviour. That is, the
height-height correlation function, $G(r-r^\prime,t) = \langle [h(r,t)
- h(r^\prime,t)]^2 \rangle^\frac{1}{2}$,
scales with time, $t$, and separation, $\ell = |r-r^\prime|$, as
\begin{equation}
\label{eqn:corr}
G(\ell,t)
\sim \ell^{\alpha}\, f(t/\ell^z).
\end{equation}
$h(r,t)$\/ is the height of the surface at position $r$\/ and
at time $t$. The dynamical scaling behaviour is characterised by the
roughness exponent, $\alpha$, and the dynamical exponent, $\beta$,
with $z = \alpha/\beta$. The scaling function $f(x)$\/ behaves as
$f(x) = x^\beta$\/ for $x \ll 1$\/ and $f(x) = $\/ {\it constant}\/
for $x \gg 1$. Thus, the surface roughness grows as $G(t) \sim
t^\beta$\/ initially, independent of size, and for a given size,
$\ell$, the roughness saturates after a sufficiently long time such
that $G(\ell)$\/ scales with $\ell$\/ only as $G(\ell) \sim
\ell^\alpha$.

Numerous simulations in a variety of growth models
\cite{Meakin86,Kosterlitz89,Family90,Tang93}
have since confirmed the hypothesis of
dynamical scaling, including models which
allow overhangs to form and models where overhangs are not allowed. An
overhang is formed when a particle sticks at
a position higher than the height of the surface at that
point, such the space below the particle is not filled.
Simulations of the restricted
solid on solid model\cite{Forrest90,Kosterlitz93} where
incoming particles fall directly onto the surface such that no
overhangs can form, and may only stick at a site if the resulting
nearest
neighbour height differences are less than
some predetermined value, have
led to a further consensus that the
values of the scaling corresponds to
that of the
Kardar-Parisi-Zhang equation \cite{kpz},
\[
\frac{\partial h}{\partial t} = \nu\, \nabla^2 h + \frac{\lambda}{2}\,
(\nabla h)^2 + \eta,
\]
where $\eta$\/ is a random variable.
This equation  is
believed to be a continuum description of deposition growth,
 and was derived by assuming that the
surface grows uniformly in the direction of the local normal.
The exponents obtained are exact in $2$\/ dimensions
\cite{kpz}, and numerically determined in higher dimensions
\cite{Amar90,Moser91}.

Results from simulations of the ballistic deposition model\cite{Vold}
where incoming particles
stick at the first point of contact and thus allow overhangs are more
controversial. At
present there is no clear consensus as to whether or not this system
belongs to the same universality class as that described by the
Kardar-Parisi-Zhang equation\cite{Banavar}, or whether the presence
of overhangs leads to a different set of scaling exponents. Early
results by Meakin {\it et. al.}\/ gave $\alpha = 0.47$ and $\beta =
0.331$
in $2$\/ dimensions, and $\alpha = 0.33$ and $\beta = 0.24$
in $3$\/  dimensions, in agreement with Kim and Kosterlitz's
approximate formula \cite{Kosterlitz89} of $\alpha = 2/(d+2)$\/ and
$\beta = 1/(d+1)$\/ for the
Kardar-Parisi-Zhang equation. More recent
results suggest that
the values of the scaling exponents may, in fact, be different. Baiod
{\it et. al.}\/ \cite{Baiod88} obtained $\beta = 1/3$\/ in $2$\/
dimensions
and $\alpha = 0.3$\/ and $\beta = 0.22$\/ in
$3$\/  dimensions; off-lattice simulations
have also given $\beta = 0.343$ in $2$\/
dimensions\cite{Jullien1}, but
a clear scaling behaviour was not observed in $3$\/  dimensions
\cite{Jullien2}.

In this Letter, we report results of simulations of ballistic
deposition and restricted solid on solid growth. We find that the
values of the scaling exponents for the
ballistic deposition model are different
to those of the restricted solid on solid model.
A summary of our results is
given in Table 1.

TABLE 1. Scaling exponents obtained from our simulations.
\begin{center}
\begin{tabular}{crrrrrr} \hline \hline
&\multicolumn{3}{c}{Ballistic}&\multicolumn{3}{c}
{Restricted}\\
dimension&\multicolumn{3}{c}{deposition}&\multicolumn{3}{c}{solid on
solid }\\
\multicolumn{1}{c}{$\qquad d \qquad$}
&\multicolumn{1}{c}{$\quad\alpha\quad$}
&\multicolumn{1}{c}{$\quad\beta\quad$} &\multicolumn{1}{c}{$\quad
z\quad$} &\multicolumn{1}{c}{$\quad\alpha\quad$}
&\multicolumn{1}{c}{$\quad\beta\quad$} &\multicolumn{1}{c}{$\quad z
\quad$} \\ \hline
$2$ & 0.45 & 0.32 & 1.40 & 0.50 & 0.33 & 1.50\\
$3$ & 0.26 & 0.21 & 1.24 & 0.40 & 0.25 & 1.60\\
$4$ & $\sim$\/ 0.12 & ---  & ---  & 0.29 & 0.18 & 1.61\\
 \hline \hline
\end{tabular}
\end{center}
\vspace{0.5cm}

\begin{figure}[t]
\vspace*{6cm}
\includegraphics{2d.g.t.ps}
\includegraphics{2d.g.L.ps}
\caption{Ballistic deposition for 2 dimensions. The inset shows
the plot of $\ln G(\ell,t)$\/ versus $\log_2 \ell$\/ at the end of the
simulation.}
\label{fig:2d.G.t}
\end{figure}

We also find that while on-lattice simulations give
excellent scaling behaviour for the restricted solid on solid model,
the same is not true for ballistic deposition. Quasi
off-lattice simulations were therefore carried out for the ballistic
deposition model. Namely, each axis of a surface of size $L^{d-1}$\/
particle diameters is
divided into $nL$\/ points such that incoming particles can
be centred on any one of these points. For $n$\/ large, the
surface approaches a continuum, and for $n = 1$, we recover
the on-lattice model.
The height of the surface at a
position $r$\/ is defined to be the height of a new particle if it
fell onto the surface at $r$. We found that $n = 3$\/ is sufficient to
give a good scaling behaviour, and no differences were found in the
results with $n = $\/ 5, 7 and 10. We performed simulations in $2$,
$3$\/  and $4$\/  dimensions for both models and the
simulations are run until the equivalent of at least 2000 layers of
atoms have been deposited. The total number of
particles deposited in each simulation is over $2 \times 10^{9}$.
The minimum time required for each run is twenty-four CPU
hours on a DEC Alpha 400 workstation. To obtain good statistics,
averages over many runs were often needed.

In figure~\ref{fig:2d.G.t}, the correlation function, $G(\ell,t)$, for
a $2$\/ dimensional ballistic deposition simulation
is plotted versus time in a log-log plot. The largest system size
considered is $\ell = 2^{20}$. For the larger values of $\ell$,
the roughness has not saturated
within the time scale of the simulation.
In the dynamical scaling region, we see a clear power law behaviour,
$G(t) \sim t^\beta$.
Also shown in the inset is
a plot of $\ln G(\ell,t)$\/ versus $\log_2 \ell$\/ for the data
at the end of our simulation.
For the smaller sizes where saturation has
been reached,
we also find a roughly linear dependence of $\ln G$\/
on $\log_2 \ell$ in agreement with the predictions of dynamical
scaling.

Direct extrapolation of the scaling exponents from the
gradients in the log-log plots turned out to be difficult because
cross-over effects due to the transition from the dynamical scaling
regime to the saturated scaling regime introduce significant
corrections.
Instead, by rewriting equation~(\ref{eqn:corr}) as
\begin{equation}
\label{eqn:scaling}
\ln G(\ell,t) - \beta \ln t = F(\alpha \ln \ell - \beta \ln t),
\end{equation}
we can obtained good estimates of the dynamical and roughness
exponents by collapsing our data for all sizes and all times
considered. This, in
fact, provides a way of checking also
whether the data corresponds to just one scaling regime, or whether
there is also a cross-over between different universality classes with
different scaling exponents. We note that the surface roughness during
the initial few time steps is strongly influenced by transient effects,
and have been discarded in the data collapse.

The collapsed data for the $2$\/
dimensional ballistic deposition result is shown in
figure~\ref{fig:2d.collapsed}. Data for $\ell$\/ ranging from $2^2$\/
to $2^{19}$\/ are used in the plot, with
over $6 \times 10^{9}$\/ particles deposited. The values
of the exponents used are $\alpha = 0.45$\/ and $\beta = 0.32$.
We have also carried out simulations of the restricted solid on solid
model in $2$\/ dimensions, and found that $\alpha = 0.50 $\/ and
$\beta
= 0.33$, in agreement with the results of previous simulations.

\begin{figure}[h]
\vspace*{6cm}
\includegraphics{2d.collapsed.new.ps}
\includegraphics{2d.collapsed.bad.ps}
\caption{Collapsed data for $2$\/ dimensional ballistic deposition
simulation, with $\alpha = 0.45$\/ and $\beta = 0.32$. The data used
ranges from $\ell = 2^2$\/ to $2^{19}$. The inset show
the collapse obtained with the exponents obtained for the restricted
solid on solid model.}
\label{fig:2d.collapsed}
\end{figure}

The collapsed data for the $3$\/  dimensional simulations
are shown in
figure~\ref{fig:3d.collapsed}. The upper diagram is for the restricted
solid on solid model. The size of the system considered is $2^{10}
\times 2^{10}$, and over $2 \times 10^{9}$\/ particles were deposited
in a run. The data presented represents the average over seven
independent runs, and include values for $\ell$\/ ranging from $2^2$\/
to $2^{9}$. The values of the scaling exponents obtained in this
case are $\alpha = 0.40$\/ and $\beta = 0.25$. This is in agreement
with the approximate formula of Kim and Kosterlitz
\cite{Kosterlitz89}, but the value of $\beta$\/ obtained is greater by
0.01 than that observed more recently by Ala-Nissila {\it et. al.}
\cite{Kosterlitz93}.

Ballistic deposition simulations in $3$\/  dimensions are also
carried
out for systems with size equal to $2^{10} \times 2^{10}$ particle
diameters, with three subdivisions per particle diameter. The
collapsed data are shown in the lower diagram of
figure~\ref{fig:3d.collapsed}.
Again, over $2 \times 10^{9}$\/ particles were deposited per run, and
the results presented represent the average over ten runs with data
for $\ell = 2^2$\/ to $2^9$\/ used in the data collapse. The
simulations were also carried out with seven subdivisions per particle
diameter and no difference was found. The value of the scaling
exponents in this case are $\alpha = 0.26$\/ and $\beta = 0.21$,
significantly lower than the corresponding values for the
restricted solid on solid model.

\begin{figure}[b]
\vspace*{12.5cm}
\includegraphics{3d.rsos.collapsed.ps}
\includegraphics{3d.collapsed.ps}
\includegraphics{3d.collapsed.bad.ps}
\caption{Collapsed data for the $3$\/  dimensional restricted
solid on
solid simulation (upper diagram), and the ballistic deposition
simulation (lower diagram) for $\ell = 2^2$\/ to $2^9$. The inset in
the lower diagram shows the
collapse obtained if the exponents obtained from the restricted solid
on solid model were used instead.}
\label{fig:3d.collapsed}
\end{figure}

We have also carried out simulations in higher dimensions for both
models. However, due to computational difficulties, we are restricted
to relatively small sizes.
For the ballistic deposition model the largest size possible in
$4$\/
dimensions or higher is still too small for
the dynamical scaling regime to be observed. Estimates of the
roughness exponent in $4$\/  dimensions, however, give a value of
$\alpha \approx 0.12$. The uncertainty in this case is due to strong
fluctuations in the roughness as a result of the small system
size, and the number of runs required to obtain better statistics is
prohibitively large.
For the restricted solid on solid model, the fluctuations
are smaller even in $4$\/  dimensions and we have been able to
obtain
reliable values for both the dynamical and the roughness exponents.
These are $\alpha = 0.29$\/ and $\beta = 0.18$, in
good agreement with those obtained by Ala-Nissila {\it et. al.}
\cite{Kosterlitz93}. Again, in accordance with the trend observed in
lower dimensions, the exponent of the ballistic
deposition model is lower than that of the restricted solid on solid
model.

\begin{figure}[b]
\vspace*{6cm}
\includegraphics{slice2d.mid.ps}
\caption{A cross-section of ballistic deposition growth in $3$\/
dimensions taken at a height of 200 particle diameters.
The length of the horizontal and vertical axis
correspond to 100 particle diameters.}
\label{fig:slice2d.mid}
\end{figure}

We have found that variations in the value of
either $\alpha$\/ or $\beta$\/ by as little as $0.01$\/ is sufficient
to give clear deterioration of the data collapse plots. The values we
present are therefore accurate to the figures quoted.
The most important implication of this is that from our results, the
dynamical scaling behaviour of the ballistic deposition model and the
restricted solid on solid model belong to different universality
classes.
We have shown in the inset to figures~\ref{fig:2d.collapsed}
and~\ref{fig:3d.collapsed} what happens when we try to collapse the
ballistic deposition data with the exponents obtained from the
corresponding restricted solid on solid simulations. It is clear from
the diagrams that even in $2$\/ dimensions where the differences
between the values of the scaling exponents for the ballistic
deposition model and those of the restricted solid on solid model are
apparently small, a satisfactory data collapse cannot be obtained.
In view of the belief that the dynamics of the restricted solid on
solid model corresponds to that of the Kardar-Parisi-Zhang equation,
our results would therefore further suggest that the
Kardar-Parisi-Zhang is not appropriate in describing deposition growth
in situations where overhangs are dominant. Indeed, our values of the
scaling exponents for the ballistic deposition model in $2$\/,
$3$\/
and in $4$\/  dimensions lie outside the range of the values for
the
Kardar-Parisi-Zhang exponents \cite{kpz,Amar90,Moser91}.

\begin{figure}[b]
\vspace*{6cm}
\includegraphics{x.ds.ps}
\caption{The fraction $x$\/ of sites occupied along a substrate
dimension versus the substrate dimension.}
\label{fig:x.ds}
\end{figure}

We have also tried to examine the structure of the solid formed by
growth under ballistic deposition conditions.
Figure~\ref{fig:slice2d.mid} show a cross-section of the bulk formed
in a $3$\/  dimensional ballistic deposition simulations. The
cross-section corresponds to a height of 200 particle diameters from
the substrate, and is taken after all the particles at this height are
covered. The cross section shown corresponds to an area of
$100 \times 100$\/ particle diameters.
We find that there are very few connected lines, and no connected
rings in the cross section.
In addition, we have calculated the fraction of sites, $x$, which are
occupied in a linear direction from the average density, $\rho$. For a
$d_s$\/ dimensional surface, the density is given by $\rho = x^{d_s}$.
In figure~\ref{fig:x.ds} a plot of $x$\/ versus substrate dimension is
shown. The results indicate that of order 0.4 of the sites along a
line on the surface are occupied in all dimensions.
This, together with the cross section plot, corroborates with
the idea that particles grow on the edges of overhangs, and almost
immediately branch off to form a complex tree like structure.

In summary, we have found that the presence of overhangs is an
important factor in determining the scaling properties of deposition
growth.  In a model such as ballistic deposition, overhangs will form
when the local surface gradient exceeds a critical value corresponding
to the presence of a sharp step in the surface profile.
In such a situation, the next particle will stick to increase the
lateral size of the overhang region rather than to reduce the surface
gradient by falling to the lower surface. Thus, as overhangs begin to
form, they will tend to increase the lateral correlation at a fast
rate,
and the surface will no longer grow in the direction of its local
gradient. The result may be an anisotropic growth which when coarse
grained lead to broader and flatter structures.
Although such a picture can give a behaviour consistent with
the results of our simulations, the search for a proper theory for
deposition growth in the presence of overhangs remain an important
challenge.

We thank H. Aoki, B.K. Chakrabarti, R.A. Cowley, K. Inata, A. Maritan,
A.L. Stella for
useful discussions. F.S. acknowledges the European Community for a
postdoctoral fellowship under the Human Capital and Mobility
Programme. This was was carried out under grant GR/G02727
of the Science and Engineering Research Council of the United Kingdom.


\begin{references}
\bibitem{Viscek.and.Family}  {\it Dynamics of Fractal Surfaces},
edited by F. Family and T. Vicsek, World Scientific, Singapore (1991);
C. Godr\`eche (ed), {\it Solids far from equilibrium},
proceedings of the Beg-Rohu Summer School (1989), Cambridge University
Press, Cambridge (1991).
\bibitem{interest} {\it See for example}:  J. Chevrier, V. le Thanh, R.
Buys and J. Darrien, Europhys. Lett. {\bf 16}, 732 (1991);
Y.L. He, H.N. Yang, T.M. Lu and G.C. Wang, Phys. Rev. Lett. {\bf 69},
3770 (1992); M.A. Cotta, R.A. Hamm, T.W. Statley, S.N.G. Chu, L.R.
Harriot, M.B. Panish and H. Temkin, Phys. Rev. Lett {\bf 70}, 4106
(1993).
\bibitem{Family} F. Family and T. Vicsek, J. Phys. {\bf A18}, L75
(1985).
\bibitem{Meakin86} P. Meakin, P. Ramanlal, L. M. Sander and R.C. Ball,
Phys. Rev {\bf A34}, 5091 (1986).
\bibitem{Kosterlitz89} J.M. Kim and J.M. Kosterlitz, Phys. Rev. Lett.
{\bf 62}, 2289 (1989)
\bibitem{Family90} F. Family, Physica {\bf A168}, 561 (1990).
\bibitem{Tang93} C. Tang and S. Liang, Phys. Rev. Lett. {\bf 71 },
2769  (1993).
\bibitem{Amar90} J. G. Amar and F. Family, Phys. Rev. A{\bf 41}, 3399
(1990).
\bibitem{Moser91} K. Moser, J. Kert\'esz and D.E. Wolf, Physica A {\bf
178}, 215 (1991).
\bibitem{Forrest90} B.M. Forrest and L.H. Tang, Phys. Rev. Lett. {\bf
64}, 1405 (1990).
\bibitem{Kosterlitz93} T. Ala--Nissila, T. Hjelt, J.M. Kosterlitz and
O. Ven\"{a}l\"{a}inen, J. Stat. Phys. {\bf 72}, 207 (1993).
\bibitem{kpz} M. Kardar, G. Parisi and Y.C. Zhang, Phys. Rev. Lett.
{\bf 56}, 889 (1986).
\bibitem{Vold} M.J. Vold, J. Colloid. Interface Sci. {\bf 14}, 168
(1959) and M.J. Vold, J. Chem. Phys. {\bf 63}, 1608 (1960).
\bibitem{Banavar} A. Maritan, F. Toigo, J. Koplik and J.R. Banavar,
Phys. Rev. Lett. {\bf 69}, 3193 (1992).
\bibitem{Baiod88} R. Baiod, D. Kessler, P. Rammanal, L. Sanderand and
R. Savit, Phys. Rev. {\bf A38}, 3672 (1988).
\bibitem{Jullien1} P. Meakin and R. Jullien, J. Physique {\bf 48},
1651 (1987).
\bibitem{Jullien2} R. Jullien and P. Meakin, Europhys. Lett. {\bf 4},
1385 (1987).
\end{references}
\end{document}